\RequirePackage{amsmath}
\documentclass[12pt]{iopart}

\usepackage{bm}
\usepackage{graphicx}
\usepackage{cite}
\usepackage{color}
\usepackage{amssymb}
\usepackage{units}

\newcommand{\vc}[1]{\bm{#1}}
\renewcommand{\r}{{\vc r}}
\newcommand{\ez}{\widehat{\vc{e}}_z}
\newcommand{\erh}{\widehat{\vc{\rho}}}
\newcommand{\eph}{\widehat{\vc{\varphi}}}

\makeatletter
\def\imod#1{\allowbreak\mkern10mu({\operator@font mod}\,\,#1)}
\makeatother

\begin{document}

\title[Spin dynamics in helical molecules with non-linear interactions]%
{Spin dynamics in helical molecules with non-linear interactions}

\author{E. D\'{i}az$^{1}$, P.Albares$^{2}$, P. G.~Est\'{e}vez$^{2}$, J. M.~Cerver\'{o}$^{2}$, C. Gaul$^{1,3}$, E.~Diez$^{2}$ and F. Dom\'{i}nguez-Adame$^{1}$}

\address{$^{1}$\ GISC, Departamento de F\'{\i}sica de Materiales, Universidad Complutense, E--28040 Madrid, Spain}

\address{$^{2}$\ NANOLAB, Departamento de F\'{\i}sica Fundamental, Universidad de Salamanca, E--37008 Salamanca, Spain}

\address{$^{3}$\ Cognitec Systems GmbH, Gro{\ss}enhainer Str.\ 101, 01127 Dresden, Germany}

\ead{elenadg@ucm.es}

\date{\today}

\begin{abstract}

It is widely admitted that the helical conformation of certain chiral molecules may induce a sizable spin selectivity observed in experiments. Spin selectivity arises as a result of the interplay between a helicity-induced spin-orbit coupling and electric dipole fields in the molecule. From the theoretical point of view, different phenomena might affect the spin dynamics in helical molecules, such as quantum dephasing, dissipation and the role of metallic contacts. Previous studies neglected the local deformation of the molecule about the carrier thus far, but this assumption seems unrealistic to describe charge transport in molecular systems. We introduce an effective model describing the electron spin dynamics in a deformable helical molecule with weak spin-orbit coupling. We find that the electron-lattice interaction allows the formation of stable solitons such as bright solitons with well defined spin projection onto the molecule axis. We present a thorough study of these bright solitons and analyze their possible impact on the spin dynamics in deformable helical molecules.

\end{abstract}


\vspace{2pc}

\maketitle

\section{Introduction} \label{sec:intro}

Manipulation and control of the electron spin degree of freedom in nanoscale materials lies at the very core of spintronics. Among the large variety of materials with technological interest in this field, organic systems are gaining significance as active components in spintronics nanodevices. Although large spin-orbit coupling~(SOC) is uncommon in carbon-based materials, recent experiments on electron transport have brought with them a considerable effort to uncover the origin of the observed high spin selectivity in DNA~\cite{Goehler11,Xie11} and bacteriorhodopsin on non-magnetic metallic substrates~\cite{Mishra11}. As a working hypothesis, it has been suggested that spin selectivity may be related to the specific geometric structure of the involved molecular systems, namely their helical conformation~\cite{Xie11}. A number of theoretical models have been put forward to explain the observed spin selectivity in helical molecules. Usually they rely on large SOC~\cite{Yeganeh09,Medina12,Gutierrez12,Eremenko13,Rai13,Medina15,Caetano16,Diaz17}, the need for dephasing when SOC is weak~\cite{Guo12,Guo14b}, the leakage of electrons from the molecule to the environment~\cite{Matityahu16}, the role of the bonding of the molecule to the metallic leads that enhance the effect~\cite{Guo14a,Wu17}, or the interplay between a helicity-induced SOC and a strong dipole electric field, which is characteristic of these molecules~\cite{Michaeli15} (see Refs.~\cite{Naaman12,Michaeli17} for a recent review). Theoretical models usually assume rigid lattices and neglect the local deformation of the molecule about the carrier. However, this assumption seems unrealistic to describe charge transport in molecular systems like DNA~\cite{Chakraborty07}.

Depending on the various energy scales involved (electron bandwidth, zero-point energy of molecular vibrations, thermal energy), lattice deformation can play a significant role on transport properties. This is parti\-cu\-larly relevant when charge carriers interact with intramolecular modes that occur at high frequency due to the stretching of stiff covalent bonds. Coupling to those modes may strongly alter charge transport~\cite{Nan09} and even lead to self-trapping of carriers, provided that the relaxation energy (the energy gained upon the deformation of the lattice about the carrier) exceeds the band width~\cite{Fratini16}. Self-trapping has been commonly formulated within the framework of the small polaron theory based on a local Holstein-type coupling~\cite{Holstein59} between the carrier and the intramolecular mode. This model was later extended by Peyrard and Bishop to study the ac response of a DNA molecule, where the charge in the $\pi$-stack interacts with the base-pair opening dynamics of the double strand~\cite{Peyrard89,Komineas02,Maniadis05,Diaz08}. 

Davydov's soliton theory of charge and energy transfer in $\alpha$-helix and acetanilide provides another paradigmatic example on how the interaction of carriers and vibrational degrees of freedom can induce self-trapping phenomena~\cite{Davydov79}. Starting from a Fr\"{o}lich-like Hamiltonian~\cite{Frolich54} and assuming the adiabatic approximation, Davydov put forward a soliton theory of long-range energy transfer of excitations interacting with intramolecular vibrational modes in a quasi-one-dimensional lattice. In the adiabatic approximation, the continuous limit of the Davydov's equations reduce to the non-linear Schr\"{o}dinger (NLS) equation for the elementary excitations. 

Inspired by the success of the Peyrard-Bishop-Holstein and Davydov's approaches, in this work we introduce an effective self-focusing non-linear model describing the dynamics of a single charge carrier in the electrostatic potential due to a helical arrangement of dipoles. The proposal generalizes the linear model formerly introduced by Guti\'{e}rrez \emph{et al.\/} considering spin-selective transport of electrons through a helically shaped electrostatic potential~\cite{Gutierrez12}. This model has been recently revisited and extended to study the coherent spin dynamics in helical molecules~\cite{Diaz17}. The strong interaction with the lattice vibrations will be addressed by adding a non-linear term to the Schr\"{o}dinger equation within the adiabatic approximation~\cite{Datta96}. The resulting equation turns out to be integrable, thus allowing us to obtain a family of bright solitons describing the coherent spin dynamics in deformable helical molecules. 

\section{Electron spin dynamics in a rigid helical molecule} \label{sec:rigid}

Following Ref.~\cite{Diaz17}, we start out by revisiting and amending the model introduced in Ref.~\cite{Gutierrez12}. Two main factors determine the high spin selectivity found: an unconventional Rashba-like SOC, reflecting the helical symmetry of molecules, and a weakly dispersive electronic band. $\alpha$-helix proteins and other macromolecules present a net dipole moment along the helix axis due to the helical arrangement of peptide dipoles~\cite{Sengupta05}. Thus we consider the electron motion through a very helical arrangement of peptide dipoles directed along the $Z$ axis. The dipoles are located at $\r_j = j \Delta z \, \ez  + a \, \erh_j$ and their dipole moments are $\vc{d}_j = d \, {\eph}_j$. Here, we have used cylindrical coordinates with $\erh_j = (\cos \varphi_j,\sin \varphi_j,0)$, $\eph_j = (-\sin \varphi_j, \cos \varphi_j,0)$, and $\varphi_j = 2\pi j/N_d +\pi$. The orientation of the individual dipoles does not affect much the results, provided they are arranged helically~\cite{Diaz17}. Typical values are $N_d=10$ dipoles per turn in DNA, $a=\unit[0.7]{nm}$, and $b=N_d \, \Delta z= \unit[3.2]{nm}$. The total electric field on the molecule axis due to the dipoles is then found to be~\cite{Diaz17}
\begin{align}
\vc{E}(z) = \frac{1}{4\pi \epsilon_0} \sum_j \frac{\vc{d}_j}{\big[a^2+(z-j\, \Delta z)^2\big]^{3/2}} \ .
\label{eq:02}
\end{align}
To estimate the SOC, we need $\mathcal{E}(z) = -i [E_x(z)-i E_y(z)] = \exp(-i 2\pi z/b)\,\mathcal{D}(z)$, where
$\mathcal{D}(z) \approx \mathcal{E}_0 \equiv d (\epsilon_0 a b \Delta z)^{-1} K_1(2\pi a/b)$, $K_1$ being the modified Bessel function of the second kind.

The SOC Hamiltonian stems from the classical formula ${\bm\sigma} \cdot (\widehat {\vc p} \times \vc{E})$, symmetrized such that the Hamiltonian is Hermitian. Here $\bm\sigma$ is a vector whose components are the Pauli matrices $\sigma_x$, $\sigma_y$, and $\sigma_z$. For $\widehat {\vc{p}} = \widehat{p}_z\,\ez$ the SOC Hamiltonian simplifies to
\begin{equation}
\widehat H_{\mathrm{SO}} = \frac{\lambda}{2} \left[ \widehat p_z 
\begin{pmatrix} 0 & \mathcal{E}(z) \\ \mathcal{E}^*(z) & 0 \end{pmatrix}
+ 
\begin{pmatrix} 0 & \mathcal{E}(z) \\ \mathcal{E}^*(z) & 0 \end{pmatrix}
\widehat p_z \right] , 
\label{eq:04}
\end{equation}
where $\lambda = e \hbar/(2mc)^2$. The electron Hamiltonian $\widehat{\mathcal{H}}=\widehat{p}_{z}^{\,2}/2m+\widehat H_{\mathrm{SO}}$ can be cast in the form $\widehat{\mathcal{H}}=E_b\widehat{H}$ where the dimensionless Hamiltonian $\widehat{H}$ reads
\begin{subequations}
\begin{equation}
\widehat{H} =-\partial^{2}_\xi-2\pi\gamma\,\widehat{M}\ .
\smallskip
\label{eq:05a}
\end{equation}
Here we have defined $E_b=\hbar^2/2mb^2$, $\xi=z/b$, $\partial_{\xi}=\partial/\partial\xi$ and the dimensionless spin-orbit parameter $\gamma = {\hbar \lambda \mathcal{E}_0}/({2\pi b E_b})$. The matrix operator $\widehat{M}$ is given by
\begin{equation}
\widehat{M}=\begin{pmatrix}
          0	& e^{-i2\pi \xi}\\
          e^{i2\pi \xi} & 0
        \end{pmatrix}
        \begin{pmatrix}
          i\partial_\xi-\pi  & 0\\
          0 & i\partial_\xi+\pi
        \end{pmatrix}\ .
\label{eq:05b}
\end{equation}
\label{eq:05}
\end{subequations}

The dimensionless Hamiltonian~\eqref{eq:05a} is readily diagonalized and the eigenenergies are found to be
\begin{subequations}
\begin{equation}
\varepsilon_{qs} = q^2 + \pi^2 - 2\pi s\sqrt{1 + \gamma^2}\,q\ ,
\qquad 
s =\pm 1\ .
\label{eq:06a}
\end{equation}
The corresponding normalized eigenfunctions are 
\begin{align}
\vc{\chi}_{qs}(\xi)&= 
  \begin{pmatrix}
   \beta_{\uparrow}(s)\, e^{i(q-\pi)\xi} \\[2pt]
   \beta_{\downarrow}(s)\, e^{i(q+\pi)\xi}
  \end{pmatrix} \ ,
\label{eq:06b}
\end{align}
where
\begin{align}
\beta_{\uparrow}(s)&=\frac{1}{2}\,\big[(1+s)\cos \phi+(1-s)\sin\phi\big]\ ,\nonumber\\
\beta_{\downarrow}(s)&=\frac{1}{2}\,\big[(1-s)\cos \phi-(1+s)\sin\phi\big]\ ,
\label{eq:06c}
\end{align}
satisfying $\beta_{\uparrow}^2(s)+\beta_{\downarrow}^2(s)=1$, and
\begin{equation}
\tan \phi = \frac{\gamma}{1+\sqrt{1+\gamma^2}}\ .
\label{eq:06d}
\end{equation}
\label{eq:06}
\end{subequations}

Once the eigenvectors of the Hamiltonian~(\ref{eq:05a}) have been obtained, we focus on the dynamics of an electron wave packet of the form
\begin{subequations}
\begin{equation}
\vc{\chi}(\xi,t)=\sum_{s}\int_{-\infty}^{\infty}\frac{dq}{2\pi}\,C_{qs }\vc{\chi}_{qs}(\xi)\,
e^{-i\varepsilon_{qs}t}\ ,
\label{eq:07a}
\end{equation}
where time is expressed in units of $\hbar/E_b$ and
\begin{equation}
C_{qs}=\int_{-\infty}^{\infty}d\xi\,\vc{\chi}_{qs}^{\dag}(\xi)\cdot\vc{\chi}(\xi,0)\ .
\label{eq:07b}
\end{equation}
\label{eq:07}
\end{subequations}
Our magnitude of interest will be the time-dependent spin projection onto the molecule axis, also referred as helicity, which is calculated as follows
\begin{align}
\mathrm{SP}(t)&=\int_{-\infty}^{\infty}d\xi\,\vc{\chi}^{\dag}(\xi,t)\sigma_z\vc{\chi}(\xi,t)
=\int_{-\infty}^{\infty}\frac{dq}{2\pi}\,\Big[\big(\,|C_{q,+1}|^2-|C_{q,-1}|^2\big)\cos(2\phi)\nonumber \\
&+2\sin(2\phi)\mathrm{Re}\left(C_{q,+1}^{*}C_{q,-1}
e^{i(\varepsilon_{q,+1}-\varepsilon_{q,-1})t}\right)\Big]\ .
\label{eq:08}
\end{align}

Consider an initial wave packet of dimensionless width $W$ with an arbitrary state of spin polarization $\vc{\chi}(\xi,0)=f(\xi)\big[\cos(\theta)\vc{u}_{\uparrow}+e^{i\varphi}\sin(\theta)\vc{u}_{\downarrow}\big]$. Here $\vc{u}_\sigma$ with $\sigma=\uparrow,\downarrow$ denotes an eigenvector of $\sigma_z$ and the polarization state is defined by the angle $\theta$. For the sake of concreteness we set $\varphi=0$ hereafter. After a straightforward calculation one can obtain a closed expression for $\mathrm{SP}(t)$ that has a transient contribution which vanishes at large times $t\gg W/\big(4\pi\sqrt{1+\gamma^2}\big)$. A transient time of $t\sim \unit[40]{fs}$ is roughly estimated for a highly localized initial state with $W\sim 1$ passing through a DNA molecule with a SOC parameter of the order of $\gamma \sim 0.1$. Thus, after a quick transient state, the spin projection reaches the asymptotic value given as $\mathrm{SP}_{\infty}=\mathrm{SP}(t\to\infty)$ where
\begin{equation}
\mathrm{SP}_{\infty}=\frac{1}{1+\gamma^2}\ \Big[ \cos(2\theta)
-\gamma \sin (2\theta) \int_{-\infty}^{\infty}d\xi\,|f(\xi)|^2 \cos(2\pi\xi)\Big]\ .
\label{eq:09}
\end{equation}
Notice that if we consider an initial \emph{fully polarized} state with $\theta=0$ or $\theta=\pi/2$, namely with spin parallel (antiparallel) to the molecule axis, the larger the SOC parameter, the smaller the asymptotic spin polarization, as expected.

In experiments, however, an initially unpolarized current becomes spin polarized after being transmitted through the helical molecule. Therefore, our case of interest is an initial \emph{fully unpolarized\/} wave packet with spin projection out of the molecule axis, i.e. along the $X$ axis such as $\theta=\pi/4$ or $\theta=3\pi/4$ and $\vc{\chi}(\xi,0)=f(\xi)\left(-\vc{u}_\uparrow+\vc{u}_\downarrow\right)/\sqrt{2}$. In such a case, the asymptotic polarization has a non-monotonous dependence with the magnitude of the SOC parameter according to Eq~(\ref{eq:10}). The integral in equation~(\ref{eq:09}) approaches unity for a narrow wave packet, and  the asymptotic spin projection along the molecule axis becomes
\begin{equation}
|\mathrm{SP}_{\infty}|=\frac{\gamma}{1+\gamma^2}\ .
\label{eq:10}
\end{equation}
Therefore, the SOC flips the electron spin after a quick transient and the spin projection along the molecule axis becomes nonzero.

\section{Electron spin dynamics in a deformable helical molecule} \label{sec:deformable}

In order to describe a deformable helical molecule where the electron dynamics is affected by the lattice vibrations, we will add a non-linear term to the Hamiltonian~(\ref{eq:05a}). This additional term can be justified within the adiabatic approximation, according to Davydov's theory~\cite{Davydov79}. In such a scenario, the dimensionless NLS describing the dynamics of the spinor state $\vc{\chi}(\xi,t)$ reads
\begin{equation}
i \partial_t \vc{\chi}(\xi,t)=\widehat{H}\vc{\chi}(\xi,t)
-4g\left[\vc{\chi}^{\dag}(\xi,t)\cdot \vc{\chi}(\xi,t)\right]\vc{\chi}(\xi,t)\ ,
\label{eq:11}
\end{equation}
where $\widehat{H}$ is given in equation~(\ref{eq:05a}).

The integrability of this equation can be analyzed by using the Painlev\'e test \cite{Estevez16}. This test proves the integrability of equation~(\ref{eq:11}) and yields its three component Lax pair. The Painlev\'e property can be also used  to derive Darboux transformations and an iterative procedure for obtaining solutions~\cite{Albares17}. It can be also proved that equation~(\ref{eq:11}) is the only integrable case of a model very recently put forward by Kartashov and Konotov to study the dynamics of Bose-Einstein condensates with helical SOC~\cite{Kartashov17}. Furthermore, it reduces to the Manakov non-linear system when the SOC vanishes~\cite{Manakov74,Vishnu13}.

 In this regard, for self-focusing non-linear interaction ($g>0$), it can be demonstrated that the following bright solitons (similar to the case of Davydov's soliton) are a solution to equation~(\ref{eq:11})
\begin{align}
\vc{\chi}_{s}(\xi,t)&= \sqrt{\frac{g}{2}}\,\mathrm{sech}\big[g(\xi +c t)\big]\,e^{-i\varphi_s(\xi,t)}
  \begin{pmatrix}
   \beta_{\uparrow}(s)\,e^{-i\pi\xi}  \\[2pt]
   \beta_{\downarrow}(s)\,e^{i\pi\xi}
  \end{pmatrix} \ ,
  \qquad
  s = \pm 1\ ,
\label{eq:12}
\end{align}
where $\beta_{\uparrow}(s)$ and $\beta_{\downarrow}(s)$ are given by~(\ref{eq:06c}). Here $c$ is a free parameter representing the velocity of the soliton. The phase is defined as $\varphi_s(\xi,t)= \left[c/2-s \pi\cos^{-1}(2\phi)\right]\xi+\left(c^2/4-g^2-\pi^2\gamma\right)t$. The existence of two different bright solitons due to the arbitrary choice of the constant $s=\pm 1$ is known as the Kramer doublet and it is directly related to the preservation of the time-reversal symmetry in the model. Notice that these solitons have a well defined state of spin polarization that depends on the SOC due to equation~(\ref{eq:06d}). Most importantly, this polarization is preserved along its propagation and it is found to be
\begin{equation}
|\mathrm{SP}_{\mathrm{sol}}|= |\beta_{\uparrow}^2(s)-\beta_{\downarrow}^2(s)| = \frac{1}{\sqrt{1+\gamma^2}}\ . 
\label{eq:13}
\end{equation}

\section{Connection with experiments} \label{sec:experiments}

Having presented the salient features of solitons in deformable helical molecules, we now turn to discuss their relevance in experiments. In recent experiments on electron transport in organic helical molecules, it has been clearly demonstrated that an initially unpolarized current turns out to be highly polarized when passing through the molecule~\cite{Goehler11,Xie11,Mishra11,Dor13,Kettner15,Mondal15,Einati15,Kiran16,Aragones17}. Since the intrinsic SOC effects are rather weak in these molecules, theoretical models proposed to describe the experiments rely on SOC related to the peculiarities of the helical geometry. It is worth mentioning that all these models strongly depend on a phenomenological SOC which was roughly estimated to be $\alpha=4-12\,$meV\,nm~\cite{Gutierrez12}. It seems that this coupling, even in the best scenario, is not large enough to support the high degree of spin polarization observed in the experiments. All these approaches, however, neglected lattice deformations that strongly affect the electron dynamics in organic molecules. 

\begin{figure}[tb]
\centerline{\includegraphics[width=0.60\linewidth]{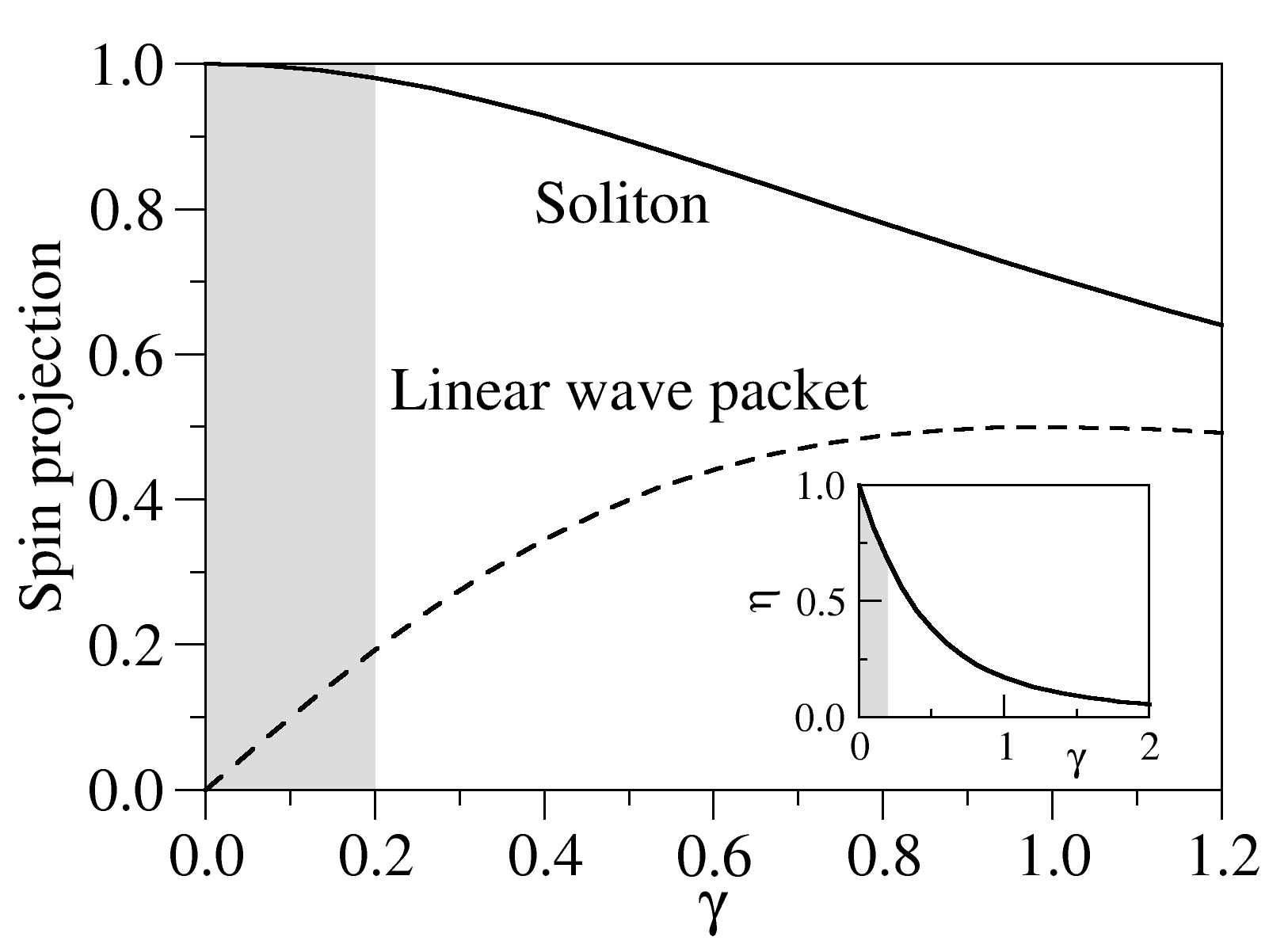}}
\caption{Asymptotic spin projection as a function of the dimensionless SOC parameter $\gamma$ for a rigid helical molecule (dashed line) and a non-linear deformable helical molecule (solid line). The inset shows the {enhancement factor\/} $\eta=\big(|\mathrm{SP}_\mathrm{sol}|-|\mathrm{SP}_\infty|\big)/\big(|\mathrm{SP}_\mathrm{sol}| + |\mathrm{SP}_\infty|\big)$. The gray area highlights the region with  realistic values of $\gamma$ according to our estimations.}
\label{fig:01}
\end{figure}

In order to show the relevance of the non-linear interaction between the lattice and the spin degree of freedom, figure~\ref{fig:01} compares the spin projection achieved with the linear, $\mathrm{SP}_\infty$, and the non-linear model, $\mathrm{SP}_\mathrm{sol}$, as given by equations~(\ref{eq:10}) and~(\ref{eq:13}) respectively. To understand the experimental situation when a spin unpolarized current is injected into the molecule, we will consider an initial localized wave packet with a polarization state such that $\theta=3\pi/4$ (fully unpolarized in the sense discussed above). The wave packet evolves in time and, after a short transient time, it reaches a steady polarization given by equation.~(\ref{eq:10}). This polarization has to be compared with that obtained for the stable soliton~(\ref{eq:13}). Figure~\ref{fig:01} clearly shows an outstanding result, namely, the non-linearity strongly enhances the resulting spin projection of a coherent electron passing through a deformable helical molecule. Thus, lattice vibrations lead to a larger effective SOC in the molecule. We define the \emph{enhancement factor\/} as $\eta = \big(|\mathrm{SP}_\mathrm{sol}| - |\mathrm{SP}_\infty|\big) /\big(|\mathrm{SP}_\mathrm{sol}| + |\mathrm{SP}_\infty|\big)$, as depicted in the inset of figure~\ref{fig:01}. This parameter assess the effects of the lattice on the spin polarization capability of the deformable helical molecule. According to the curve shown in the inset of figure~\ref{fig:01}, the self-focusing non-linear interaction has a major effect when the SOC is weak. And this is precisely the case of interest in experiments, as discussed below. 

The gray area of figure~\ref{fig:01} highlights the region where the dimensionless SOC of our model takes values consistent with previous estimations $\gamma=\alpha/(2\pi b E_b) \leq 0.2$ in DNA. Our approach explains the physical scenario in experiments as follows. The initial unpolarized electron is injected in a deformable helical molecule whose vibrations interact continuously with the electronic dynamics inside the system. Such non-linear interaction transforms the initial arbitrary state till it reaches the most stable configuration, namely, a solitonic solution. Once the soliton state is conformed, it can propagate with no dispersion along the molecule with a well defined high degree of spin polarization. The formation of solitons after a short transient time could be an important effect for the resulting highly polarized currents found in experiments.

\section{Conclusions}

In conclusion, we have presented a non-linear model to study the spin dynamics of electrons in a deformable helical molecule. Such dynamics is subjected to:  i) the electric field created by the helical arrangement of molecular dipoles and ii) the interaction between the electron and the lattice vibrations. On the one hand, the dipole electric field induces a Rashba-like SOC for electrons moving along the helical axis. On the other hand, the electron-lattice interaction allows the formation of stable solitons.

Once the model was presented, we were able to prove that the system supports the formation of stable bright solitons.
Remarkably, the spin polarization in such solitons is preserved during the propagation across the helical molecule. We also calculated the spin projection onto the molecule axis as a figure of merit to assess the spin dynamics.  For completeness, we also compared this result with those obtained for the linear model corresponding to the rigid molecule. In particular, we focused on the most relevant situation for experiments, namely the partial spin polarization achieved by an initial unpolarized electron when passing through the helical molecule. In such scenario, the spin polarization capability of a deformable helical molecule is largely increased compared to the rigid one, even in the case of weak SOC. 

\ack

The authors thank R.~Guti\'{e}rrez and V. Mujica for helpful discussions. This research has been supported by MINECO (Grants MAT2013-46308 and MAT2016-75955) and Junta de Castilla y Le\'{o}n (Grant SA045U16).

\section*{References}

\bibliography{references}

\bibliographystyle{iopart-num.bst}

\end{document}